\documentclass[twocolumn,aps,prc]{revtex4}
\usepackage{epsfig}
\newcommand{\beq}{\begin{equation}}
\newcommand{\eeq}{\end{equation}}
\newcommand{\bea}{\vspace{0.25cm}\begin{eqnarray}}
\newcommand{\eea}{\end{eqnarray}}

\newcommand{\pb}{{{\bf p}}}

\def\lsim{\mathrel{\rlap{\lower4pt\hbox{\hskip1pt$\sim$}}
    \raise1pt\hbox{$<$}}}         
\def\gsim{\mathrel{\rlap{\lower4pt\hbox{\hskip1pt$\sim$}}
    \raise1pt\hbox{$>$}}}         
\begin{document}

\renewcommand{\thefootnote}{\fnsymbol{footnote}}

\title{\Large\bf
Medium modification of photon-tagged and inclusive jets in high-multiplicity 
proton-proton collisions
}
\renewcommand{\thefootnote}{\arabic{footnote}}

\newcommand{\landau}{L.D.~Landau Institute for Theoretical Physics,
        GSP-1, 117940, Kosygina Str. 2, 117334 Moscow, Russia}

\author{B.G.~Zakharov} \affiliation{\landau}


\begin{abstract}
We study modification of the photon-tagged and inclusive 
jets in $pp$ collisions at $\sqrt{s}=7$ TeV 
due to mini-quark-gluon plasma which can be produced in high multiplicity 
events. We show that for underlying events with 
$dN_{ch}/d\eta\sim 20-60$
the medium effects lead to a considerable modification of 
the photon-tagged and inclusive jet fragmentation functions. 
For inclusive jets the magnitude of the effect is 
surprisingly large. 
The effect is quite strong even for typical 
underlying events. 
We find that 
the spectrum of charged hadrons is suppressed 
by $\sim 35-40$\% at $p_{T}\sim 5-10$ GeV.
\end{abstract}


\maketitle


The experiments at RHIC and LHC
have provided strong evidence for the formation 
in $AA$ collisions of a hot and dense QCD matter in the form of 
quark-gluon plasma (QGP).
However, so far we have no compelling evidence for the formation of 
the QGP in  $pp$ collisions.
It is widely believed that
over the energy range covered by the accelerator experiments
in the typical inelastic minimum bias $pp$ collisions 
the collectivity in the final states 
is unimportant due to small energy density. 
Nevertheless, in rare high-multiplicity (HM) $pp$ events the energy density
may be comparable to that in $AA$ collisions
at RHIC and LHC energies. And if the thermalization time, $\tau_{0}$, 
is small enough,
say $\tau_{0}\lsim 0.5$ fm, the QGP should be formed quite likely to 
$AA$ collisions.

One can expect that
collectivity in HM $pp$ collisions may be observed
via the hydrodynamic flow 
effects \cite{Bozek_pp,Wied_pp} similar to that in $AA$ collisions.
The hydrodynamical simulation of the 
flow effects in $pp$ collisions
with the initial conditions from the IP-Glasma model \cite{IPG12} has been
performed in \cite{glasma_pp}. It was observed that 
for small-size QGP the theoretical uncertainties 
are large as compared to the QGP in $AA$ collisions.
This may make it difficult to extract information on the
mini-QGP via measurements of the azimuthal flow coefficients.

Although today we do not have 
clear evidence for the formation of the QGP in $pp$ collisions,
there are some indications 
in its favor. 
It is possible that 
observed  by the CMS collaboration 
ridge correlation structure in HM $pp$ events 
\cite{CMS_ridge1} is  
due to  the transverse flow of the mini-QGP. But alternative 
explanation of this effect in the color glass condensate picture seems also 
possible \cite{CGC_ridge}.
In \cite{Camp1},
employing Van Hove's idea \cite{VH} that  
phase transition should  lead to anomalous behavior of 
$\langle p_{T}\rangle$ as a function multiplicity, 
it has been argued that the $pp$ data on 
$\langle p_{T} \rangle$ 
signal possible plasma formation in the domain $dN_{ch}/d\eta\sim  6-24$.

An unambiguous proof of formation of a dense QCD matter 
in $pp$ collisions would be observation of jet 
modification (quenching) similar to that observed in $AA$ collisions.
It is important that conditions for the QGP production are better
in events with jets, because multiplicity of soft off-jet particles (so-called 
underlying events (UEs)) is enhanced by a factor of $2-3$ \cite{CDF}.
In $AA$ collisions 
jet quenching leads to
suppression of high-$p_{T}$ 
spectra
characterized by the nuclear
modification factor $R_{AA}$ defined as the ratio
of the particle yield in the $AA$ collision to the binary-scaled
yield in $pp$ collisions. The latter
provides the baseline which for now is assumed to be free of 
the final state medium effects.
However, for $pp$ collisions the medium modification factor, 
$R_{pp}$, is unobservable directly
because we do not have the baseline spectra 
with switched off the final state interactions in the QGP. 
For observation of the medium effects in $pp$ collisions  
measurement of the jet fragmentation function (FF)
in $\gamma+$jet events seems to be promising, as was suggested 
in \cite{W12} for $AA$ collisions.
To understand the prospects 
of this method 
in this Letter we evaluate the medium modification
of the $\gamma$-tagged FF at $\sqrt{s}=7$ TeV in the midrapidity region 
near $y=0$ for different multiplicities of the UE.
We also give predictions for the medium modification of 
the FF for inclusive jets, which is closely related to 
$R_{pp}$.
Although $R_{pp}$ is unobservable directly it is important
for the nuclear modification factors
$R_{pA}$ and $R_{AA}$,
which, in the presence of the mini-QGP, 
should be divided by $R_{pp}$. 
To illustrate the magnitude of the suppression effect in $pp$ collisions 
we present our preliminary results 
for $R_{pp}$ of charged hadrons.

The jet quenching 
is dominated by radiative 
energy loss \cite{BDMPS,LCPI,BSZ,GLV1,AMY}  with relatively small effect from 
collisional mechanism \cite{BSZ,Z_Ecoll,Gale}. 
We use 
the light-cone path integral (LCPI) approach \cite{LCPI,BSZ} to induced 
gluon emission. 
We treat the effect of parton energy loss on 
the FFs within  
the scheme developed previously for $AA$ collisions \cite{RAA08}.
It takes into account both radiative and
collisional energy loss. 
This approach was successful in explaining results for  
$R_{AA}$ for light and heavy flavors in $AA$ collisions 
\cite{RAA11,RAA12,RAA13}.
We do not discuss the details of the model, and 
refer the reader to our above cited articles
on jet quenching in $AA$ collisions.

As in \cite{RAA08} we use the 1+1D Bjorken's model of the
QGP evolution, which gives $T_{0}^{3}\tau_{0}=T^{3}\tau$, 
and take $\tau_{0}=0.5$ fm. 
For $\tau<\tau_{0}$ we take medium density $\propto \tau$. 
However, the effect of this region is relatively small.
We neglect variation of the initial temperature $T_{0}$ with the 
transverse coordinates.
To fix $T_{0}$ 
we use the entropy/multiplicity ratio 
$c=dS/dy{\Big/}dN_{ch}/d\eta\approx 7.67$ obtained in \cite{BM-entropy}.
The initial entropy density can be written as 
\beq
s_{0}=\frac{c}{\tau_{0}\pi R_{f}^{2}}\frac{dN_{ch}}{d\eta}\,,
\label{eq:10}
\eeq
where $R_{f}$ is the radius of the created fireball
(we assume that the jet production is dominated by the head-on collisions
and ignore azimuthal asymmetry of the QGP).
One can expect that 
$R_{f}\sim R_{p}\sim 1$ fm (here $R_{p}$ is the proton radius).
It agrees qualitatively with $R_{f}$ obtained 
for $pp$ collisions at $\sqrt{s}=7$ TeV in numerical simulations
performed in \cite{glasma_pp} 
within the IP-Glasma model \cite{IPG12}. The $R_{f}$ from \cite{glasma_pp}
grows approximately as linear function of $(dN_{g}/dy)^{1/3}$ and then 
flattens. 
We use the $R_{f}$ from \cite{glasma_pp} 
parametrized in a convenient form via $dN_{g}/dy$ in \cite{RPP}.
The values of the $R_{f}$ and $T_{0}$ for 
different values of 
$dN_{ch}/d\eta$ (we take $dN_{g}/dy\approx 2.13 dN_{ch}/d\eta$)
obtained using the ideal gas model with $N_{f}=2.5$
are given in Table I.
One sees that for $dN_{ch}/d\eta\gsim 20$ (\ref{eq:10}) gives $T_{0}$
well above the deconfinement temperature $T_{c}\approx 170$ MeV.
For $dN_{ch}/d\eta=40$ we have $T_{0}$ close to that for the central
Au+Au collisions at $\sqrt{s}=200$ GeV \cite{RAA13}.
\vspace{-.5cm} 
\begin{table}[h]
\caption{$R_{f}$ and $T_{0}$ for different $dN_{ch}/d\eta$.}
\begin{tabular}{|c|c|c|c|c|c|} 
\hline
$dN_{ch}/d\eta$ & 3 & 6 & 20 & 40 & 60 \\
\hline
$R_{f}$ (fm) & 1.046 & 1.27 & 1.538 & 1.538  & 1.538 \\
\hline
$T_{0}$ (MeV) & 177 & 196 & 258 & 325 & 372 \\
\hline
\end{tabular}
\end{table}

In $\gamma$+jet events
the energy of the produced hard parton, $E_{T}$, in the direction opposite 
to the tagged direct photon 
is smeared around the photon energy, $E_{T}^{\gamma}$. 
 The NLO calculations
\cite{Wang_NLO2} show that for $AA$ collisions 
at $E_{T}^{\gamma}\sim 8$
the smearing  correction,
$\Delta_{sm}$, to the medium modification factor $I_{AA}(z)$ of the
photon tagged FF 
blows up at $z\gsim 0.8-0.9$ (hereafter $z=p_{T}^{h}/E_{T}^{\gamma}$). 
One can show that 
$\Delta_{sm}\approx F(z,E_{T}^{\gamma})dI_{AA}/dz/E_{T}^{\gamma\,\,2}$,
where $F(z,E_{T}^{\gamma})$ is a smooth function of $E_{T}^{\gamma}$.
Using this formula 
and the results of \cite{Wang_NLO2}  (shown in Fig.~2)
one can show that for $E_{T}^{\gamma}\gsim 25$ GeV 
(which will
be considered in our Letter) the effect of smearing should be very small at
$z\lsim 0.9$.
To be conservative we will consider the region $z<0.8$, where the effect of
smearing is practically negligible and the LO relation $E_{T}=E_{T}^{\gamma}$
can be used.
Then 
following \cite{W12} 
we can write the $\gamma$-tagged FF in $pp$ collisions as a 
function of the UE multiplicity density $dN_{ch}/d\eta$ (for
clarity we denote it by $m$) as
\beq
D_{h}(z,E_{T}^{\gamma},m)\!=\!\big\langle\!\big\langle
\!\sum_{i} r_{i}(E_{T}^{\gamma})D_{h/i}(z,E_{T}^{\gamma},m)
\!\big\rangle\!\big\rangle,
\label{eq:20}
\eeq
where
$D_{h/i}$ is the medium modified FF for $i\to h$ process, and 
$r_{i}$ is the fraction of the $\gamma+i$ parton state in the 
$\gamma+$jet events, $\langle\!\langle...\rangle\!\rangle$ means averaging
over the transverse geometrical variables of $pp$ collision 
and jet production, which includes averaging over the fast parton path length
$L$ in the QGP.

For not very high $E_{T}^{\gamma}$ the sum over
all relevant types of partons on the right-hand side of (\ref{eq:20}) 
is dominated by gluon and light quarks. 
For all light quarks
medium modification of the FFs are very similar, 
and we may consider one effective light quark state  $q$ with
$r_{q}=1-r_{g}$. 
We use for the $r_i$ prediction of the ordinary LO, the pQCD
formula, which gives $r_g\ll r_q$ at LHC
energies.
In principle, $r_g/r_q$ may depend on $m$.
However, there are
no serious physical reasons for the strong multiplicity dependence of
this ratio, because the UE activity is driven by fluctuations
of soft gluons which should not strongly  modify  the hard cross sections.
And since the dominating
contribution to the $\gamma$-tagged jets 
comes from quark jets, the theoretical uncertainties due to 
modification of the $r_g/r_q$ ratio should not be significant.

We have performed averaging over $L$ 
using the distribution of hard processes in the impact parameter plane 
obtained with the quark distribution from the MIT bag model.
It is plausible for our preliminary study. In the MIT bag model 
practically in the full range of the impact parameter the distribution
in $L$ is sharply peaked around $L\approx\sqrt{S_{ov}/\pi}$, where $S_{ov}$
is the overlap area for two colliding bags. It means that the effective 
fireball radius $R_{f}$ (which includes all centralities) at the same
time gives the typical path length for fast partons.
We have found that the effect of the $L$ fluctuations
is relatively small (as compared to $L=R_{f}$ they
reduce the medium modification by $\sim 10-15$\%).

As in \cite{RAA08,RAA11,RAA12,RAA13} 
we evaluate radiative and collisional energy 
loss with running $\alpha_s$
frozen at some value $\alpha_{s}^{fr}$ at low momenta.
For gluon emission in vacuum a reasonable choice is $\alpha_{s}^{fr}\approx 0.7$
\cite{NZ_HERA}. But in plasma thermal effects can suppress
$\alpha_{s}^{fr}$.
We observed previously \cite{RAA13} that data on $R_{AA}$ 
are consistent with $\alpha_{s}^{fr}\approx 0.5$ for RHIC
and $\alpha_{s}^{fr}\approx 0.4$ for LHC.
The reduction of $\alpha_{s}^{fr}$ from RHIC to LHC 
may be related to stronger thermal effects in the QGP due to 
higher initial temperature at LHC. As noted, for the mini-QGP
produced in $pp$ collisions the gluon formation 
length is of the order of or larger than the medium size. In this
regime in the LCPI treatment \cite{LCPI} a large contribution to 
the induced gluon spectrum 
comes from configurations with interference of the emission amplitude 
and the complex conjugate one 
when one of them typically has the gluon emission point outside the medium.
For this reason for $pp$ collisions 
$\alpha_{s}^{fr}$ may be somewhat larger than that obtained for $AA$
collisions (for same $T_{0}$).
We take $\alpha_{s}^{fr}=0.6$. The results are not very sensitive
to variation of $\alpha_{s}^{fr}$ in the physically reasonable domain  
$\alpha_{s}^{fr}\sim 0.5-0.7$. The point is that  
for a small-size plasma the hardness $Q$ of induced gluon emission can 
attain quite large
values since $Q^{2}\gsim 2\omega/L$ \cite{Z_Ecoll}. 
For radiation of gluons with energy $\omega\sim 1-3$ GeV and $L\sim 1$ fm 
$Q\gsim 0.6-1$ GeV, where the freezing of $\alpha_{s}$ is not very 
important.
\begin{figure} [t]
\vspace{.7cm}
\begin{center}
\epsfig{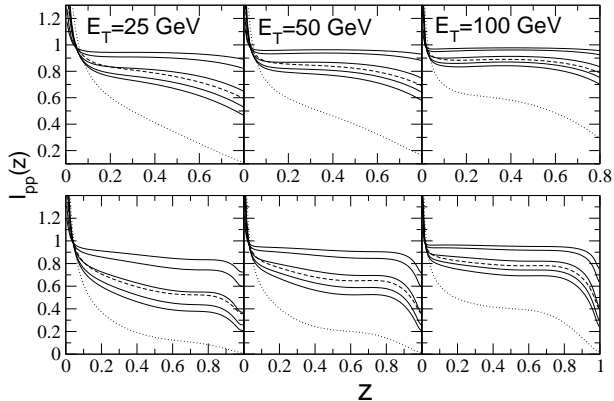}
\end{center}
\vspace{-0.5cm}
\caption[.]
{$I_{pp}$ for $\gamma$-tagged
(upper panels) and inclusive (lower panels)  jet FFs  
at $\sqrt{s}=7$ TeV for $dN_{ch}/d\eta=[3, 6, 20, 40, 60]$ 
(solid line). The order (top to 
bottom) of the curves
at large $z$ corresponds to increasing values of
$dN_{ch}/d\eta$.
The dashed line shows ratio of the FFs for
$dN_{ch}/d\eta=40$ and $3$. The dotted line shows the medium
modification factor at $\sqrt{s}=2.76$ TeV for the QGP 
with $T_{0}=420$ MeV and $L=5$ fm for $\alpha_{s}^{fr}=0.4$.
}
\end{figure}

In Fig.~1 we present the results for the medium modification factor
(for charged hadrons)
\beq
I_{pp}(z,E_{T},m)=D_{h}(z,E_{T},m)/D^{vac}_{h}(z,E_{T})\,
\label{eq:30}
\eeq
for the $\gamma$-tagged (upper panels) and inclusive (lower panels) jets
for  $E_{T}=[25, 50, 100]$ GeV at $\sqrt{s}=7$ TeV
(as in (\ref{eq:20}) $m=dN_{ch}/d\eta$). We used the LO pQCD predictions: 
$r_g/r_q\approx [0.093, 0.12,0.17]$ for $\gamma$-tagged jets and
$r_g/r_q\approx [6.99, 5.68, 4.25]$ for inclusive jets for our set of $E_{T}$.
The smearing effect is irrelevant to inclusive jets, and we show
the results for the whole range of $z$ since  it is interesting in the context
of the suppression factor of the high-$p_{T}$ spectra.
In principle, our treatment of multiple 
gluon emission, based on Landau's method
\cite{BDMS_RAA}, is not supposed to be valid at very small $z$,
where cascading of the primary gluons radiated from the fast partons
comes into play. We included the region of very small $z$ just to 
illustrate the flow of jet energy into the soft region.
For illustration of the difference between 
the medium effects in $pp$ and $AA$ collisions we also present the curves for 
$\sqrt{s}=2.76$ TeV  for $L=5$ fm and $T_{0}=420$ MeV that can be regarded 
as reasonable values for Pb+Pb collisions (we used  $\alpha_{s}^{fr}=0.4$, 
which is favored by the data on $R_{AA}(p_{T})$ at $p_{T}\gsim 20$ GeV).
Figure 1 shows that there is a considerable quenching effect for 
$dN_{ch/d\eta}\gsim 20$. The effect is stronger for inclusive jets
since for the $\gamma$-tagged jets the dominating contribution comes from
quarks and for inclusive ones from gluons.
In practice, for the $\gamma$-tagged jets one should simply compare the FFs
for different multiplicities (since the vacuum FF is unobservable). 
For illustration, we show the ratio of the FFs for $m=40$ and $m=3$
(for inclusive jets this ratio cannot be measured, we show it just 
to illustrate the difference between the $\gamma$-tagged and inclusive
jets).
One 
sees that for the observation
of jet quenching at $E_{T}\sim 25-50$ GeV  it is necessary to measure 
the $\gamma$-tagged FF with rather small errors, 
say,  smaller than 10\% for the UE with 
$dN_{ch}/d\eta\sim 40$ at $z\sim 0.5-0.8$.

To estimate the 
errors
related to uncertainty in the fireball size
we have performed the calculations for $R_{f}$ increased by a factor of  $1.3$. 
We have found a very small variation of $I_{pp}$, typically $|\Delta
I_{pp}/(1-I_{pp})|\lsim 0.01-0.05$. 
And even for $R_{f}$ increased 
by a factor of $1.5$ the variation remains approximately at the same level.
This says that the $I_{pp}$, regarded as a functional of the density profile
along the jet trajectory, is quite stable against  variations of this profile
(for a fixed initial entropy). 
This is due to a strong compensation between the enhancement 
of the energy loss caused by increase of the medium size and its 
suppression due to reduction of the medium density.
This test also provides a strong argument that the 
transverse expansion, neglected in our analysis, should not 
dramatically modify  our results. 
Indeed, from the point of view of the
induced gluon emission there is nothing special in the variation 
of the density profile generated by the transverse expansion. And the 
magnitude of the hydrodynamic variation of the density profile is of the order 
of that in our test. 
For jet quenching in $AA$ collisions the smallness of the hydrodynamical 
effects was demonstrated in \cite{BMS_hydro}.
In $pp$ collisions their role should be reduced
since the typical formation length for induced gluon emission is of 
the order of the medium size or larger. In this regime parton 
energy loss is mostly controlled by 
the mean amount of the matter traversed by fast partons, and the details of 
the density profile are not very important.

The observed strong quenching of inclusive jets is 
qualitatively supported by the preliminary data from 
ALICE \cite{ALICE_jet_UE} indicating
that for the HM UE jets undergo a softer fragmentation.
Note that even for typical UE
when $dN_{ch}/d\eta\sim 14$ \cite{ATLAS_UE_Nch} at $\sqrt{s}=7$ TeV, 
at moderate $z$ the suppression
is $\sim 20-30$\%. It seems to be in contradiction with 
the jet FF measured
in Pb+Pb collisions at $\sqrt{s}=2.76$ TeV \cite{CMS_jetFF} which gives
$I_{AA}$ close to unity. However, 
in Ref. \cite{CMS_jetFF} 
$z$ is defined 
through the 
energy inside the jet
cone, which should be smaller than the energy of the primary 
parton
due to the energy of soft partons deposited in the plasma or outside the jet
cone. The data on $R_{AA}$ 
indicate that the real jet FF
is strongly suppressed. At $\sqrt{s}=2.76$ TeV the suppression should be
like that shown by the dotted curve in Fig.~1.  

The medium modification factor for hadron spectra 
can be written through the medium modified FFs 
in the form
\bea
R_{pp}(p_T)=\frac{\sum_{i}\int_{0}^{1} \frac{dz}{z^{2}}
D_{h/i}(z,p_{T}^{i})
\frac{d\sigma(pp\rightarrow iX)}{d\pb_{T}^{i} dy}}
{
\sum_{i}\int_{0}^{1} \frac{dz}{z^{2}}
D_{h/i}^{vac}(z, p_{T}^{i})
\frac{d\sigma(pp\to iX)}{d\pb_{T}^{i} dy}
}\,,
\label{eq:90}
\eea
where
${d\sigma(pp\to iX)}/{d\pb_{T}^{i} dy}$ 
is the ordinary hard cross section,
$\pb_{T}^{i}=\pb_{T}/z$ is the parton 
transverse momentum. 
In (\ref{eq:90}) it is implicit that $D_{h/i}$ 
is averaged over the jet production point,
the impact parameter of  $pp$ collision and the UE multiplicity.
Equation (\ref{eq:90}) can be thought of as an analogue of the formula
for $R_{AA}$ in the whole impact parameter range. 
Presently we do not have information about the 
UE activity for each impact parameter and transverse position of the 
jet production. We evaluated the medium modified FFs 
$D_{h/i}$ averaging over $L$ but using simply the average 
UE multiplicity measured by ATLAS
\cite{ATLAS_UE_Nch}.
\begin{figure} [t]
\vspace{.7cm}
\begin{center}
\epsfig{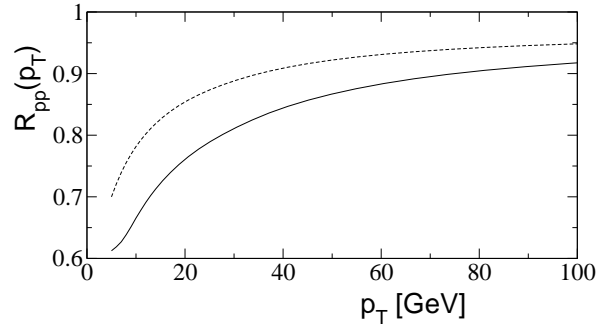}
\end{center}
\vspace{-0.5cm}
\caption[.]
{$R_{pp}$ of charged hadrons at 
$\sqrt{s}=7$ TeV for the parameters of the fireball obtained with 
the UE (solid line) and minimum bias (dashed line) $dN_{ch}/d\eta$.
}
\end{figure}
The $dN_{ch}/d\eta$ from Ref. \cite{ATLAS_UE_Nch} grows with momentum of the 
leading charged jet hadron at $p_{T}^{l}\lsim 3-5$ GeV and then flattens 
 at $dN_{ch}/d\eta\approx 13.9$
(it corresponds to $R_{f}\approx 1.51$ fm and 
$T_{0}\approx 232$ MeV).
Simulation with PYTHIA \cite{PYTHIA} shows that for jets with 
energy $E\lsim 15$ GeV, that can feel the jet energy dependence 
of the UE multiplicity, one can simply 
take $p_{T}^{l}=\eta p_{T}$, where $p_{T}$ is the hadron momentum
in Eq. (\ref{eq:90}) and $\eta\sim 1.9$ for LHC energies.
Figure 2 shows $R_{pp}$ at $\sqrt{s}=7$ TeV 
obtained with $\alpha_{s}^{fr}=0.6$.
The suppression is quite strong:
$\sim 35-40$\% at $p_{T}\sim 5-10$ GeV.
We also show $R_{pp}$ for the minimum bias multiplicity 
$dN_{ch}^{mb}/d\eta=6.01$ 
\cite{ALICE_mb_dNch}.
Even in this case the effect is significant.
The medium suppression should be important for $pA$ collisions as well.
The data on $R_{pPb}$ at $\sqrt{s}=5.02$ TeV  from ALICE \cite{RpPb}
show a small deviation from unity at $p_{T}\gsim 10$ GeV, where the Cronin
effect is weak. In the light of the strong medium suppression 
for $pp$ collisions, this may be consistent with the scenario with the
mini-QGP only if the UE multiplicities for $pp$ and $pPb$ collisions
are close to each other.
The detailed results on the $R_{pp}$
and its effect on $R_{AA}$ and  $R_{pA}$ will be presented elsewhere.

In summary, assuming that a mini-QGP fireball may be created in
$pp$ collisions, we have evaluated 
the medium modification
factors for the $\gamma$-tagged and inclusive jet FFs for $\sqrt{s}=7$ TeV.
We show that in $pp$ collisions
with UE multiplicity density $dN_{ch}/d\eta\sim 20-40$
the mini-QGP  can suppress
the $\gamma$-tagged FF at $E_{T}\sim 25-100$ GeV and $z\sim 0.5-0.8$
by $\sim 10-40$\%, and for inclusive jets the effect is even stronger.
The formation of the mini-QGP also leads to a sizeable 
suppression of the high-$p_{T}$ spectra in $pp$ collisions
$R_{pp}\sim 0.6-0.65$ at $p_{T}\sim 5-10$ GeV for 
$\sqrt{s}=7$ TeV. 
The deviation of $R_{pp}$ from unity will increase the theoretical 
predictions for the $R_{AA}$ and $R_{pA}$.
Because of a smaller suppression of the heavy flavors
the effect of the mini-QGP in $pp$ collisions 
may be important for the jet flavor tomography of $AA$ \cite{BG,RAA13}.

\vspace{-.5cm}
\begin{acknowledgments}
I am grateful to I.P.~Lokhtin and D.V.~Perepelitsa
for useful information.
This work is supported 
in part by Grant 
No.  RFBR
12-02-00063-a.
\end{acknowledgments}

\end{document}